\documentclass[nofootinbib,notitlepage,onecolumn,secnumarabic,amssymb, nobibnotes, aps, prd]{revtex4-1}
\usepackage{graphicx}
\pdfoutput=1 
\usepackage{amsmath}
\usepackage{cancel}
\usepackage{bm}
\usepackage[utf8]{inputenc}
\usepackage{multirow}
\usepackage{amsfonts}
\usepackage{xcolor}
\usepackage{hyperref}
\usepackage{epstopdf}
\usepackage{natbib}
\def\d {{d}}

\def\k {\bm{k}}
\def\q {\bm{q}}
\def\x {\bm{x}}

\newcommand{\ba}{\begin{eqnarray}}
\newcommand{\ea}{\end{eqnarray}}

\newcommand{\HH}{\mathcal{H}}

\newcommand{\two}{^{\text{\tiny ({{2}})}}}
\newcommand{\one}{^{\text{\tiny ({{1}})}}}

\def\jcap{JCAP}

\DeclareUnicodeCharacter{2212}{-}
\DeclareUnicodeCharacter{2212}{-}

\begin{document}
\title{General relativistic  effects in the galaxy bias at second order}
\author{Obinna Umeh$^{a,b}$, Kazuya Koyama$^{a}$, Roy Maartens$^{a,b}$, Fabian Schmidt$^{c}$, Chris Clarkson$^{d,b,e}$\\
\emph{\normalsize 
$^a$Institute of Cosmology \& Gravitation, University of Portsmouth, Portsmouth PO1 3FX, United Kingdom\\
$^b$Department of Physics \& Astronomy, University of the Western Cape,Cape Town 7535, South Africa \\
 $^{c}$Max-Planck-Institut f\"{u}r Astrophysik, Karl-Schwarzschild-Str. 1, 85741 Garching, Germany\\
{{$^{d}$School of Physics \& Astronomy, Queen Mary University of London, London E1 4NS, UK\\
$^e$Department of Mathematics and Applied Mathematics, University of Cape Town, Rondebosch 7701, South Africa}}
}}

\date{\today}

\begin{abstract}

The local galaxy bias formalism relies on the energy constraint equation at the formation time to relate the metric perturbation to the matter density contrast.  In the Newtonian {approximation}, this relationship is linear, which allows us to specify the initial galaxy density as a function of local physical operators. In {general} relativity however, the relationship is intrinsically nonlinear and a modulation of the short-wavelength mode by the long-wavelength mode {might be} expected.  We describe in detail how to obtain local coordinates where the coupling of the long- to the short-wavelength {modes}  is removed through a change of coordinates {(in the absence of primordial non-Gaussianity)}. We derive the general-relativistic correction to the galaxy bias expansion at second order. The correction does not come from the modulation of small-scale clustering by the long-wavelength mode; instead, it arises from distortions of the volume element by the long-wavelength mode and it does not lead to new bias parameters.

\end{abstract}

\maketitle
 \setcounter{footnote}{0}
\DeclareGraphicsRule{.wmf}{bmp}{jpg}{}{}
\maketitle

\section{Introduction}

{Next-generation large-scale stucture (LSS) surveys such as  Euclid~\cite{Amendola:2016saw}, LSST~\cite{Zhan:2017uwu} and the SKA~\cite{Maartens:2015mra}, together with cross-correlations between these surveys~\cite{Fonseca:2015laa,Alonso:2015sfa}, will probe the distribution of galaxies on ultra-large scales (above the equality scale), where the effects of general relativity (GR) can become important.} It is therefore imperative that the theoretical model for these tracers is formulated consistently in GR.  

We do not yet {fully} understand how galaxies form from the initial curvature perturbation and then evolve under a given theory of gravity to become what we observe today,  forcing us to adopt an effective field theory-like approach to model their large-scale behaviour. This approach uses perturbation theory techniques to model  tracers as a function of the long-wavelength mode of a set of physical operators, while averaging over the short-wavelength mode component within a local patch. The averaged contribution of the short modes is then incorporated as bias parameters, which may be determined from observations or  N-body simulations. The goal here is to be able to describe the observed statistics of any tracer with as many bias parameters as may be required within a range of scales where the perturbation theory description may be trusted. The bias parameters appear as coefficients of the physical operators $\mathcal{O}(\tau, {\x})$ in a perturbation theory expansion of the tracer proper number density contrast
\begin{eqnarray}\label{eq:generalbiaspres}
\delta_{\rm{g} }(\tau ,{\x})=\sum_{n}b_{\mathcal{O}^{n}}(\tau) \{\mathcal{O}(\tau ,{\x})\}^{n}\,,
\end{eqnarray}
where $\delta_{\rm{g}} = n_{\rm{g}}/\bar{{n}}_{\rm{g} }-1$ is the density contrast of a particular galaxy type, $b_{\mathcal{O}}$ is the bias parameter, $n_{\rm{g}}$ is the galaxy number density and $\bar{n}_{\rm{g}}$  is the mean. {$\mathcal{O}(\tau ,{\x})$  consists of a set of operators that may be constructed from the irreducible decomposition of higher than one spatial derivative of the initial curvature perturbations~\cite{McDonald:2009dh}.} The crucial result is that the evolution over long time scales, which is natural in the case of LSS tracers, can be dealt with order by order in perturbation theory \cite{Senatore:2014eva,Mirbabayi:2014zca}. That is, even though the formation history is highly nonlocal in time, the bias expansion can be written as local in time, as in 
equation \eqref{eq:generalbiaspres}. Whether the expansion is performed at initial time (Lagrangian) or at final time (Eulerian) is then a matter of choice. More generally, any complete linearly independent combination of these observables at any order in perturbation theory leads to an equivalent bias expansion~\cite{McDonald:2009dh,Desjacques:2016bnm}.

An open problem in the effective description of biased tracers is to determine the number of distinctive operators that need to be included for an accurate description of galaxy clustering at any given order in perturbation theory. Within  the Newtonian approximation, it is now well understood what key operators are to be included and also how to construct them from the initial curvature perturbations \cite{Desjacques:2016bnm}. This is straightforward in the Newtonian approximation mainly because the energy constraint equation, Poisson's equation, which relates the initial curvature perturbation to the matter density contrast, is linear and there is a unique Eulerian frame in which the initial density field is related to the evolved density field. On {ultra-large} scales, we need to apply GR, and its energy constraint equation leads to a nonlinear relationship between the curvature perturbation and the matter density field. In addition, the Eulerian frame is not unique in GR, hence the gauge choice becomes an issue as well.

{{Galaxy bias in GR was first discussed in \cite{Challinor:2011bk,Bruni:2011ta,Baldauf:2011bh,Jeong:2011as} but at the linear order.  The importance of expressing the galaxy bias model in local coordinates was discussed in \cite{Baldauf:2011bh,dePutter:2015vga}. 
The  comoving-synchronous gauge  was identified as a unique gauge choice in a Lagrangian frame for specifying the initial galaxy bias in  GR at second order in \cite{Bertacca:2015mca}. The details on  how to construct a consistent local coordinate valid on horizon scale up to second order in perturbation theory were discussed in \cite{Dai:2015jaa}. There is currently no study that  brings these pieces of information together to give a  consistent expression for the local galaxy bias model in GR at second order.  This is a gap we aim to fill.}}

{The main purpose of this paper is to provide a derivation of the local galaxy bias model (including tidal stress) within GR at second order in perturbation theory in a universe dominated by dust plus a cosmological constant,
  for Gaussian initial conditions. We also investigate how GR effects influence galaxy clustering.} 

The structure of this paper is as follows. In section II, we provide a formula for the conservation of galaxy number in the comoving-synchronous gauge, which is the unique Lagrangian frame in GR \cite{Bertacca:2015mca}.
In section III, we introduce the local coordinates at initial time where the initial galaxy density is related to a set of local operators constructed from the initial curvature perturbation. Using the conservation of galaxy number,  we relate the initial galaxy density  to the evolved galaxy density. We show that distortion of the Lagrangian volume leads to a GR correction to the galaxy bias model at second order. In section IV, we show how to relate the galaxy bias model in comoving-synchronous gauge (C-gauge) to various Eulerian gauges such as the total matter (or comoving orthogonal) gauge  (T-gauge) \cite{Villa:2015ppa},  Poisson gauge (P-gauge), N-body gauge \cite{Fidler:2017pnb}, N-Boisson gauge \cite{Fidler:2018geb}, and other possible Eulerian gauges.

{\bf{Notation}:} {Greek letters denote space-time  indices and Latin denote spatial indices. The (averaged) 4-velocity field of galaxies/ matter is $u^\mu=dx^\mu/d\tau$, and the world-lines are labelled by the comoving coordinates $\q$. The matter density contrast is $\delta_{\rm m}= \delta \rho_{\rm m}/\bar{\rho}_{\rm m}$ and the galaxy number density contrast is $\delta_{\rm g}= \delta n_{\rm g}/\bar{n}_{\rm g}$.
We expand perturbations up to second order as
  $ \delta = \delta\one +  \delta\two /2$. }
  
\section{Conservation of galaxy number  in general relativity}

\begin{figure}[!h]
\includegraphics[width=80mm] {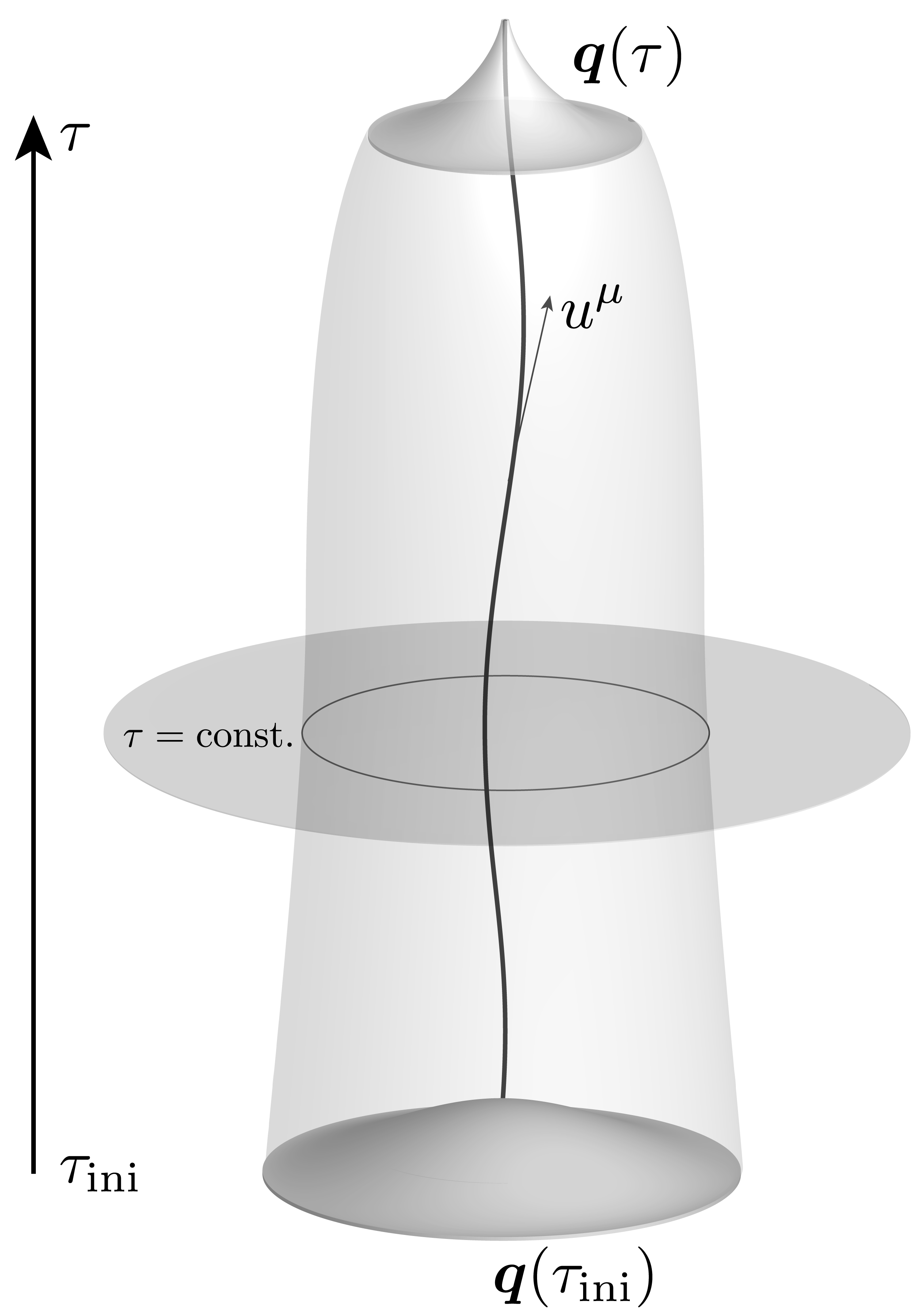}
\caption{We assume that galaxies form  around peaks (see figure \ref{fig:patches} for details)  of the matter density field {at $\tau_{\rm{ini}}$ and evolve to $\tau$ in accordance  with a set of conservation and propagation equations.  
A $\tau=\,$constant  hypersurface orthogonal to $u^{\mu}$ is shown. } }
\label{fig:Geodesiceqn}
\end{figure} 
We show in figure \ref{fig:Geodesiceqn} the conceptual set-up for the formation and  evolution of galaxies that we describe as biased tracers of the dark matter density field. In this set-up,  galaxies form within a patch of radius $R$ and evolve under the influence of gravity until observed at proper time $\tau$. 

Firstly, we derive the relation between  galaxy densities at $\tau_{\rm{ini}}$ and $\tau$ in C-gauge. The C-gauge is a unique Lagrangian frame for irrotational dust  fluid in GR. The line element in C-gauge is 
\begin{equation}
ds^2 = - d\tau^2 + a^2(\tau) \gamma_{i j}(\tau, \q) dq^i dq^j\,,
\end{equation}
where  $a(\tau)$ is the scale factor and $\gamma_{i j}$ is the conformal metric on hypersurfaces orthogonal to $u^\mu$. 
The conservation {of galaxy number  within a small volume $dV_C$ in GR} is given by, 
\begin{eqnarray}\label{eq:GRconservation}
{dN=n_{gC }(\tau,\q)dV_C(\tau,\q)=n_{gC }(\tau_{\rm{ini}},\q)dV_C(\tau_{\rm{ini}},\q)\,, \qquad dV_C=a^3\sqrt{\gamma}\,d^3q\,.}
\end{eqnarray}
Using $n_{\rm{g C}} = \bar{n}_{\rm{g}}(1 + \delta _{\rm{g C}})$, equation \eqref{eq:GRconservation} implies
\begin{equation}
1 + \delta_{{\rm{gC}}}( {\tau, \q}) = \big[1 +\delta_{\rm{g}}^{\rm{L
}}({\q}) \big]\, \frac{\sqrt{\gamma^{{\rm{ini}}}({{\q}})}}{\sqrt{\gamma(\tau,{{\q}})}}\,.
\label{eq:deltagc}
\end{equation}
Here ${\delta_{{\rm{gC}}}=\delta n_{\rm{gC}}/\bar{n}_{\rm g}}$ is the  galaxy density contrast  in C-gauge, $\gamma^{\rm{ini}}_{ij}({\q})\equiv \gamma_{ij}(\tau_{\rm{ini}},{\q})$ is the `seed' metric and  $\delta_{\rm{g}}^{\rm{L
}}({\q}) \equiv \delta_{\rm{gC }}(\tau_{\rm{ini}},{\q})$ is the initial (Lagrangian) galaxy density contrast.

The evolved metric can be written in terms of the displacement field $\beta^i$ as  \cite{Rampf:2014mga}
\begin{equation}\label{eq:metric}
 a^2(\tau)\gamma_{ij}(\tau,{{\q}}) dq^i dq^j =  a^2(\tau)
\big[1 - 2 \mathcal{B}(\tau,{{\q}}) \big]  \gamma^{ {\rm{ini}}}_{kl} ({{\q}})
\big[ \delta^k_i +\partial_i\beta^k (\tau,{{\q}}) \big]
\big[ {\delta^l_j} +\partial_j\beta^l (\tau,{{\q}}) \big] 
dq^i dq^j\,,
\end{equation}
where $ \mathcal{B}$ is the trace of the scalar perturbations of $\gamma_{ij}$~{(see Appendix \ref{sec:CosmoPert} for details)}.
In the limit {$\mathcal{B}\to 0$},  
\begin{equation}
 \frac{\sqrt{\gamma^{{\rm{ini}}}({{\q}})}}{\sqrt{\gamma(\tau,{{\q}})}}
 ~~  {\to} ~~ J (\tau, \q) \equiv \det \big[\delta^k_i +\partial_i\beta^k(\tau,{{\q}}) \big]\,,
\end{equation}
we reproduce the Newtonian {approximation
\ba
1 + \delta_{{\rm{gC}}}( {\tau, \q}) = \big[1 +\delta_{\rm{g}}^{\rm{L
}}({\q}) \big]\,J( {\tau, \q})\,\qquad (\mathcal{B}=0)\,,
\ea
to the full GR expression \eqref{eq:deltagc}.} 
The seed metric is given by \cite{Bruni:2014xma,Rampf:2014mga}
\begin{eqnarray}
\gamma_{  ij}^{\rm{ini}} = \delta_{ij}\,{\exp}\left[-\frac{10}{3} \Phi(\tau_{\rm{ini}},{{\q}})\right]=\delta_{ij}\left[1 - \frac{10}{3} \Phi(\tau_{\rm{ini}},{{\q}})+ \frac{50}{9}\big(\Phi(\tau_{\rm{ini}},{{\q}})\big)^2\right]\,,
\label{eq:initialmetric}
\end{eqnarray}
where $ {\Phi^{\rm{ini}}}({\q}) \equiv \Phi(\tau_{\rm{ini}},{\q})$ is  related to the initial curvature perturbation $\zeta = -{5} \Phi^{\rm{ini}}/3$. 
Assuming that the initial matter density field is nearly smooth~\cite{Matsubara:2008wx}, the conservation of the matter density field implies that
\begin{equation}\label{eq:ratioofdeterminants}
1+\delta_{\rm{mC}}(\tau,\q) = \frac{\sqrt{\gamma^{{\rm{ini}}}({{\q}})}}{\sqrt{\gamma(\tau,{{\q}})}} \,.
\end{equation}
Note that in C-gauge there is a residual spatial gauge freedom, $q^i \rightarrow q^i + \xi^i (q^k)$, corresponding to how one assigns coordinates $q^i$ to the 
fluid world-lines. This residual gauge does not affect 
$\sqrt{\gamma^{\rm{ini}}}/{\sqrt{\gamma} }$, thus the evolved matter density is 
invariant under this residual spatial gauge transformation.
{Using  equation \eqref{eq:ratioofdeterminants}, the GR conservation equation   \eqref{eq:deltagc} can be rewritten as}
\begin{eqnarray}\label{eq:conservationgen}
1+ \delta_{\rm{gC}}(\tau,{ \q})& =&
\left[ 1 +\delta_{\rm{g}}^{\rm{L
}}({\q})\right]\left[1 + \delta_{\rm{mC}}(\tau, {\q})\right]\,.
\end{eqnarray}
{This has the {\it same} form as the Newtonian approximation -- but the latter is arrived at by implicitly neglecting the contribution of  the metric perturbation $\mathcal{B}$.}

At linear order, the GR and Newtonian results agree: $\delta\one _{\rm{mC}} =\delta\one _{\rm{mN,C}} = \delta_{\rm{m}}\one$. 
{At second order, we can decompose $\delta_{\rm{mC}}$ into a Newtonian part and a GR correction $\delta\two_{\rm{mC}}= \delta_{\rm{mN,C}}\two +\delta_{{\rm{mGR,C}}}\two$}, where the Newtonian part is given by
\begin{eqnarray}
 {\delta_{\rm{mN,C}}\two}({\tau},\q)&=& 
 \frac{2}{3} \left[2+\frac{{ F{(\tau)}}}{{ D{(\tau)}}^2}\right] \big(\delta\one_{\rm{m}}({\tau},\q)\big)^2  + \left[1 - \frac{{F{(\tau)}}}{{D{(\tau)}}^2}\right] s^2({\tau},\q)\,.
    \label{eq:secondorderdensityCSgauge}
\end{eqnarray}
Here  $D$ is the matter growth factor at linear order and $s^2 = s_{ij}s^{ij}$, where the tidal field $s_{ij}$ is defined below.   In an Einstein-de Sitter background, ${F}(\tau)=3[{D}(\tau)]^2/7$, and this is a very good approximation in $\Lambda$CDM as well.  The second-order GR correction 
 is given by \cite{Bruni:2013qta,Villa:2015ppa}
\begin{eqnarray}
\delta\two_{{\rm{{mGR,C}}}}({\q})={ 6\Omega_{\rm m } \HH^2} \left[ -\frac{1}{4}\partial_i \nabla^{-2} \delta\one_{\rm{m}}({\q}) \partial^i \nabla^{-2} \delta\one_{\rm{m}}({\q})
    +\delta\one_{\rm{m}} ({\q})\nabla^{-2} \delta\one_{\rm{m}}({\q}) \right] \Big(1+\frac{2}{3}{f\over\Omega_{\rm{m}}}\Big)\,,
    \label{eq:GRdensity}
\end{eqnarray}
where  $\Omega_{\rm{m}}$ is the matter-energy density parameter,  $f$ is the growth rate; in a $\Lambda$CDM universe $f = \Omega_{\rm{m}}^{0.55}$ is a good approximation.   {{We have introduced the conformal Hubble parameter $\HH = \HH(\eta)$, where $\eta$ is the conformal time,  related to $\tau$ according to $d\tau = a d \eta$}}.  $\nabla^{-2} \delta\one_{\rm{m}}({\q})  \sim \Phi({\q})$ and $\partial_i \nabla^{-2} \delta\one_{\rm{m}}({\q}) \sim \partial_i v({\q})$, {with $v$ being  the peculiar velocity potential}. {We have suppressed the $\tau$ dependence for brevity.}

\subsection{ Effective theory-like description of galaxy formation}
In the previous section, we provided the relation between the initial and evolved galaxy density in C-gauge. In this section, we  provide the expression for the initial galaxy density. The idea here is to describe the complicated physics of galaxy formation using a set of bias parameters that relates the galaxy density to a set of local observables constructed from the initial curvature perturbation. The bias parameters hide our ignorance of the detailed physics of galaxy formation. The key difficulty though is how to define these local observables, since we cannot simply expand $\delta_{\rm{g}}^{\rm{L}}$ as a functional of a set of all terms that could be formed from $\Phi$, for example  $ \mathcal{F}(\Phi,\partial _i\Phi, \partial_{i}\partial_{j}\Phi)$.  The reason is that the gradient of $\Phi$ is coordinate dependent and the relationship between $\Phi$ and $\delta_{\rm{mC}}$ in GR differs from its equivalent in the Newtonian limit. We first review the Newtonian case. 

\begin{figure}[!h]
~
\includegraphics[width=140mm] {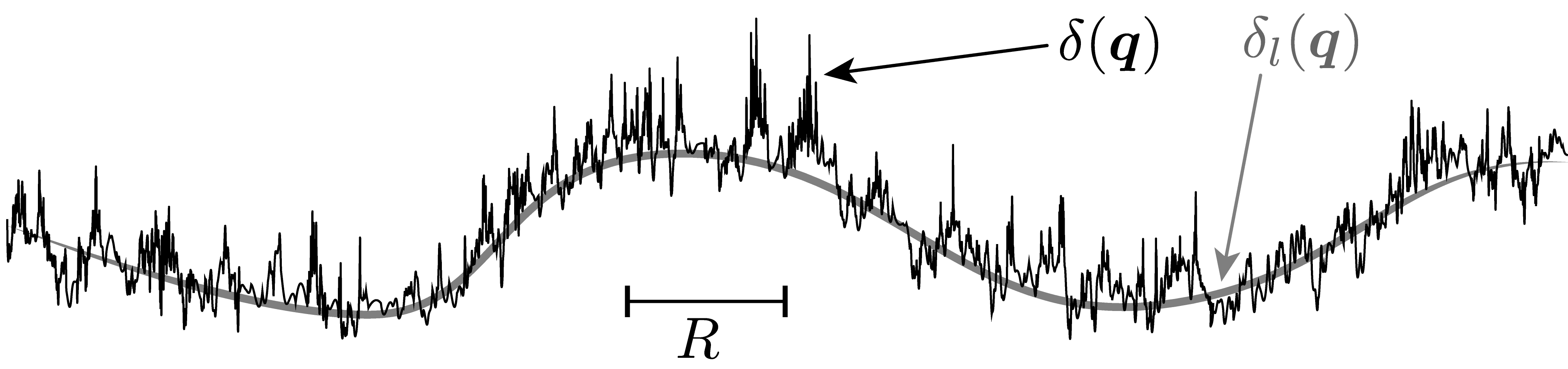}
\caption{The thick grey line is the long mode of the dark matter density field, while the snaky black line is the total dark matter density fluctuation. The short modes are responsible for the intrinsic scatter around $\delta_{\l}$.  
 }
\label{fig:patches}
\end{figure} 

\begin{itemize}

\item {\bf{Lagrangian bias in the Newtonian approximation:}}
The initial galaxy density may be expressed in terms of the  local observables  constructed from the initial metric fluctuation, $\delta_{\rm{g}}^{\rm{L}}({\q}) =\mathcal{F}[\Phi^{\rm{ini}}({\q})]$, but we need to remove unphysical modes. We make a local gradient expansion of $\Phi^{\rm{ini}}({\q})$ about a reference world-line $\bm{q} = 0$,
\begin{eqnarray}
 {\Phi^{\rm{ini}}}({\q}) &=& \Phi^{\rm{ini}}_0+( \partial_{i}{\Phi^{\rm{ini}}_0}) q^{i} 
  + \frac{1}{2}(\partial_{i}\partial_{j}{\Phi^{\rm{ini}}_0}) q^{i}q^{j}  + \mathcal{O}({\q})^{3} \,,
  \label{eq:Taylorseries}
\end{eqnarray} 
where we set $\Phi_{0}^{\rm{ini}} \equiv \Phi^{\rm{ini}}(\bm{0})$, $ \partial_{i}{\Phi^{\rm{ini}}_0} \equiv\left( \partial_{i}{\Phi^{\rm{ini}}}\right)_0$  and $\partial_{i}\partial_{j}{\Phi^{\rm{ini}}_0} \equiv \left(\partial_{i}\partial_{j}{\Phi^{\rm{ini}}}\right)_0$ to reduce clutter.
The second term does not have any influence on physics within the local patch as it can be removed by a change of coordinates,
\begin{eqnarray} \label{eq:coordaintetransform}
\tilde{\eta} = \eta\,,\qquad \qquad
\tilde{q} ^i= {q}^i + {{\xi}}^i(\eta)\,,
\end{eqnarray}
where $\xi$ is a spatially constant 3-vector. Under this coordinate change, $\Phi$ transforms as \cite{Kehagias:2013yd}
\begin{eqnarray}
{\Phi^{\rm{ini}}}(\tilde{\q}) = \Phi^{\rm{ini}}({\q}) - \left( \partial_{\eta}^{2} {\xi}_{i}+ \HH \partial_{\eta} {\xi}_{{i}}\right)  {q}^i.
\label{eq:NewtonianConfKillingeqn}
\end{eqnarray}
The constant gradient $\partial_{i}\Phi^{\rm{ini}}_0$ can be removed by demanding that  ${{\xi}^i}$ solves the equation 
\begin{equation}
\partial_{\eta}^{2} {\xi}^{i}+ \HH \partial_{\eta} {\xi}^{{i}}= \partial^{i}\Phi^{\rm{ini}}_0.
\end{equation}
In the new coordinates, the gradient term vanishes and the first term is simply a constant and may be absorbed into the scale factor of the background spacetime. 
The coordinate transformation that removes this term corresponds to  {{re-labelling of the fluid world-lines}} in the presence of the long-wavelength mode~\cite{Desjacques:2016bnm}. In the new coordinates, the galaxy position is slightly displaced, ${\q} + {\bf{\xi}} = \tilde{\q}\rightarrow {\bf{\xi}}$,  and this coordinate change leaves equation \eqref{eq:conservationgen}  invariant.
Thus, $\delta_{\rm{g}}^{\rm{L}} $ is only a functional of $\partial_{i}\partial_{j}\Phi^{\rm{ini}}_0 $ and its higher derivatives.  The observables are then constructed from the irreducible  decomposition
\begin{eqnarray}
\partial_{i}\partial_{j}\Phi^{\rm{ini}}_0  = \frac{1}{3}\nabla^2 \Phi_0^{\rm{ini}}\delta_{ij} + s_{0ij} \,,
\label{eq:irrducibledecomp}
\end{eqnarray}
where $s_{0ij}$ is the initial tidal tensor: $s_{0ij} = {D}_{ij} \Phi^{\rm{ini}}_0$ and ${D}_{ij} $ is given by ${D}_{ij} =  {\partial_{i}\partial_{j}} - {\nabla^{2}} \delta_{ij}/3$ .  Furthermore, the energy density constraint equation  in the Newtonian approximation is the Poisson equation
\begin{eqnarray}
 {\nabla^2 \Phi} (\tilde{\q}) = \frac{3}{2} {{\Omega_{\rm{m}} \HH^2}}\delta_{{\rm{mN}}}(\tilde{\q}) \,, \label{eq:firstorderdensityCSgauge}
\end{eqnarray}
which gives a linear relationship between $\Phi$ and $\delta_{\rm{mN}}$ at all orders in perturbation theory~\cite{Bernardeau:2001qr}.
By decomposing $\delta_{{\rm{mN}}}$ into long and short modes, $\delta_{{\rm{mN}}}  = \delta_s + \delta_{\l}$, it is possible to  specify $\delta_{\rm{g}}^{\rm{L}}(\tilde{\q})$ as a local functional of the long mode, $
\delta_{\rm{g}}^{\rm{L
}}(\tilde{\q}) = \mathcal{F}\left[\delta_{\l}(\tilde{\q}), s^2_{\l}(\tilde{\q})\right] \,,
$
where the effect of the short mode is encoded in the bias parameters. 
 
\item {\bf{Lagrangian bias in general relativity}:}
We now turn into the case of GR. In GR, the energy constraint equation that links $\Phi$ to $\delta_{\rm{ m}}$ is a nonlinear equation\footnote{This is derived from the Hamiltonian constraint in a $1+3$ covariant decomposition~\cite{Ellis1971grc..conf..104E,Ellis:1998ct}.}. At second order in C-gauge it is given by~\cite{Bruni:2014xma,Bartolo:2015qva}
\begin{eqnarray}
\nabla^2 \Phi ({\q}) +\bigg[2 \Phi ({\q})\nabla^2 \Phi({\q}) - \frac{1}{2}\partial_i\Phi ({\q})\partial^i \Phi({\q})\bigg]\left(1 + \frac{2}{3} \frac{f}{\Omega_{\rm{m}}}\right) = \frac{3}{2} \Omega_{\rm{m}} \HH^2 \delta_{\rm{mC}}({\q})\,.
\label{eq:GRPoissoneqn}
\end{eqnarray}
An equivalent expression exists in Possion gauge~\cite{Hidalgo:2013mba} and in total matter gauge~\cite{Villa:2015ppa}.
We perform a short/long-wavelength mode decomposition,  
neglecting gradients of the long modes relative to those of the short modes, and using $\Phi_s\nabla^2\Phi_{l} \ll \Phi_l\nabla^2 \Phi_s$. This leads to
\begin{eqnarray}
\nabla^2 \Phi_s({\q})  +\bigg[ 2 \Phi_{\l}({\q}) \nabla^2 \Phi_s ({\q})- \partial_i\Phi_{\l}({\q}) \partial^i \Phi_s({\q}) \bigg]\left(1 + \frac{2}{3} \frac{f}{\Omega_{\rm{m}}}\right)= \frac{3}{2} \Omega_{\rm{m}} \HH^2 \delta_{\rm{mC}s}({\q})\,,
\label{eq:longshortcoupling}
\end{eqnarray}
for the short mode. 
This equation may suggest that clustering on small scales is modulated by the long-wavelength mode in a similar way to primordial non-Gaussianity of the local type~\cite{Bruni:2014xma}, and this would in principle require additional bias parameters to capture these effects.

However, this is not the case as we now show.  Firstly, we have to elevate the  coordinate transformation (equation \eqref{eq:coordaintetransform}) that removes the unphysical mode in the Newtonian limit to full GR. Similar to the Newtonian approach, we consider the following coordinate transformation
\begin{eqnarray}
\tilde{q} ^{i}&= &{q}^{i} + {{\xi}}^{i}.
\label{eq:localtrajectory}
\end{eqnarray}
Under this change of coordinates the spatial metric transforms as
\begin{eqnarray}
\tilde{\gamma}_{ij} - {\gamma}_{ij} =  \xi^{k} \partial_{k} {\gamma}_{ij} + {\gamma}_{kj} \partial_{i} \xi^{k} + {\gamma}_{ik} \partial_{j} \xi^{k} \,,
\label{eq:metrictransform}
\end{eqnarray}
where the right-hand side is equal to the Lie derivative of $\gamma_{ij}$, $\mathcal{L}_{\xi} \gamma_{ij}$.
We then seek  $\xi^{i}$ that solves the following conformal Killing equation 
\begin{eqnarray}
 \mathcal{L}_{\xi} \gamma_{ij} = \left( \mathcal{A} -1\right)\gamma_{ij} \,,
 \label{eq:CKE}
\end{eqnarray}
such that the metric transforms as $\tilde{\gamma}_{ij} = \mathcal{A}{\gamma}_{ij}$, where $\mathcal{A}$ is an effective conformal factor. $\xi$ that solves equation \eqref{eq:CKE} for the Euclidean metric, $\delta_{ij}$, is given by \cite{levine1936,levine1939}
\begin{eqnarray}
{\xi}^{i} &=&  a^{i} + M^{i}{}_{j} q^{j} + \lambda q^{i}+ 2 \left( q_{j} b^{j}\right) q^{i} -  q_jq^j b^{i}\,,
\label{eq:solntoCKE}
\end{eqnarray}
where $a^{i}$, $b^{j}$ and $\lambda$ are infinitesimally small and spatially constant.  $M^{i}{}_{j} $ and $a^{i} $ are associated with the spatial rotations and translations respectively, $\lambda$ is related to the dilatation   and $b^{j}$ is related to the  special conformal transformation.  The effective conformal factor may then be expressed in terms of $\xi^{i}$ as
\begin{eqnarray}
\mathcal{A}= 1  +  \frac{2}{3}\partial^{i} \xi_{i} = 1+2\left(\lambda + 2 q^{i} b_{i}\right)\,.
\label{A}
\end{eqnarray}
Dropping the spatial rotations and translations since they leave the metric unchanged,  equation \eqref{eq:localtrajectory} becomes
\begin{eqnarray} \label{eq:etacoord2}
\tilde{\eta}&=&\eta \,,
\\
 \tilde{q}^i&=&{q}^i + \lambda q^i + 2 q^i  q^jb_j -{q}_j {q}^j b^i\,.
\label{eq:xcoord2}
\end{eqnarray}

Now applying the spatial diffeomorphism to the initial metric perturbation,
 \begin{eqnarray}
 \gamma_{ij}^{\rm{ini}} =\delta_{ij} {\exp}\left(-\frac{10}{3} \Phi^{\rm{ini}}({{\q}})\right)\,,
 \end{eqnarray}
 and using equation (\ref{A}), we can determine $\lambda$ and $b^{j}$ that remove the conformal factor of the initial metric for constant $\Phi^{\rm{ini}}_{0}$ and $\partial_i\Phi^{\rm{ini}}_{0}$. This gives the transformation  
\begin{eqnarray}
\tilde{\eta}& =& \eta \,,
\\
\tilde{q}^i &=&q^i (1- \frac{5}{3}\Phi^{\rm{ini}}_{0})  -\frac{5}{3} q^i {q}^j \partial_j \Phi^{\rm{ini}}_{0}  + \frac{5}{6} q_j q^j\partial^i \Phi^{\rm{ini}}_{0}\,.
\label{eq:xcoordbody}
\end{eqnarray}
$\Phi^{\rm{ini}}_{0}$ is related to the long-wavelength mode of the matter density perturbation \cite{Tram:2016cpy,Koyama:2018ttg}
\begin{eqnarray}
\Phi^{\rm{ini}}_{0}=\frac{9}{10} \Omega_{\rm{m}} \HH^2 \left[1+ \frac{2f}{3 \Omega_{\rm{m}}} \right] \nabla^{-2}\delta_0\,,
\label{eq:zetatodelta_L}
\end{eqnarray}
where $\delta_0$ is the leading order term in the gradient expansion of $\delta_{\l}$
\begin{equation}
\delta_l({\q}) = \delta_0+ (\partial_{i}{\delta_0}) q^{i} + \cdots. 
\end{equation}
Again we set $\delta_0\equiv \delta({\bf{0)}}$ and  $\partial_{i}{\delta_0 }\equiv (\partial_{i}{\delta})_0$.
{In Appendix~\ref{sec:CFC}, we show that this coordinate transformation is equivalent at this order to going to the conformal Fermi coordinates defined by
  \cite{Pajer:2013ana,Dai:2015rda,Dai:2015jaa}.}

Under the local coordinate transformation given in equation \eqref{eq:xcoordbody}, the linear matter density transforms as 
\begin{eqnarray}
 \delta_{\rm{mC}}(  \tilde{\q})  = \delta_{\rm{mC}} ({\q}) + \partial_j \delta_{\rm{mC}}\left({\q}\right)  \left(  \tilde{q}^j-{q}^j\right)\,.
\label{eq:CSgaugemap}
\end{eqnarray}
We split the density into short and long modes as $\delta_{\rm{mC}} = \delta_{l} + \delta_s$ and perform the transformation for the short mode density. 
After lengthy algebra (see details in {{Appendix \ref{sec:generallongmode}}}), equation \eqref{eq:CSgaugemap} becomes
\begin{eqnarray}
\delta_s( \tilde{\q}) & =&  \delta_s({\q})  
 +{3}{\Omega_{\rm{m}} \HH^2}\left[ 1 + \frac{2 f}{3 \Omega_{\rm{m}}}   \right]\delta_{{s}}({\q}) \nabla^{-2} \delta_0 
+\frac{3}{2} \Omega_{\rm{m}} \HH^2\left[1+ \frac{2f}{3 \Omega_{\rm{m}}} \right] \partial^j\nabla^{-2} \delta_{s}({\q})\partial_j  \nabla^{-2}\delta_{\rm{0}}\,.
\label{eq:densitycoord}
\end{eqnarray}
Substituting equation \eqref{eq:densitycoord} in equation \eqref{eq:longshortcoupling}  removes all the long-short coupling contribution therein. Thus, in local coordinates or within the local patch, the matter density contrast at second order in C-gauge is well approximated by the Newtonian density field and the local patch is not modulated by the long mode, hence  no new bias parameter  is required. We reach the same conclusion by constructing local coordinates using conformal Fermi coordinates (CFC) as shown in Appendix~\ref{sec:CFC}. 

\end{itemize}

We can now specify  $\delta^{\text{L}}_{\rm{g}}({\tilde{\q}}) $ as a functional of local observables constructed from the second derivatives of $\Phi_l$:
\begin{eqnarray}
1+\delta^{\text{L}}_{\rm{g}}(\tilde{\q}) = \mathcal{F}\left[\delta_{\l}(\tilde{\q}), s^2_{\l}(\tilde{\q})\right] = 1+b^\text{L}_{1}\delta_{\l}\one(\tilde{\q})
+\frac{1}{2}\left[b_{2}^\text{L}(\delta_{\l}\one(\tilde{\q}))^2+ b_{s}^\text{L}s^2_{\l}(\tilde{\q})\right]\,,
\label{eq:halooverdensity}
\end{eqnarray}
where $b^\text{L}_{1}$  and $b_{2}^\text{L}$ are linear and nonlinear Lagrangian bias parameters respectively and $b_{s}^\text{L}$ is the initial tidal bias parameter~\cite{Desjacques:2016bnm};
\begin{eqnarray}
b^\text{L}_{1} \equiv\frac{D_{\rm{ini}}}{D} \frac{\partial {\mathcal{F}}}{\partial \delta_{\l}}
\,, \qquad
b^\text{L}_{2} \equiv\left(\frac{D_{\rm{ini}}}{D}\right)^2 \frac{\partial^2 {\mathcal{F}}}{\partial \delta_{\l}^2}
\,, \qquad
b^\text{L}_{s} \equiv\left(\frac{D_{\rm{ini}}}{D}\right)^2 \frac{\partial {\mathcal{F}}}{\partial s^2_{\l}}\,.
\end{eqnarray}
The evolution of the  galaxy number is dependent on the hypersurface or the trajectory of the fluid flow, so we need to transform equation \eqref{eq:conservationgen} to the local coordinates using  equation \eqref{eq:localtrajectory};
\begin{eqnarray}
1+ \delta_{{\rm{gC}}\l}( \tilde{\q})& =&
 \left[ 1 + \delta^{\text{L}}_{\rm{g}}(\tilde{\q})\right]\left[1 + \delta_{\rm{mC}}(\tau, \tilde{\q}) \right]
\\
& =&1 + \left[1 + b_1^{\rm{L}}\right]\delta_{\l}\one(\tilde{\q})
+ \frac{1}{2} \left[\delta\two_{{\rm mN} \l} (\tilde{\q}) +\delta\two_{{\rm{mGR,C}} \l}(\tilde{\q})+  \left[b_2^{\rm{L}} +2 b_1^{\rm{L}}\right]\left(\delta_{\l}\one(\tilde{\q})\right)^2 +  b_{s}^{\rm{L}}s^2_{\l}(\tilde{\q})\right]\,.
\label{eq:galaxibias}
\end{eqnarray}
In the first line, we used the fact that the conservation equation is unchanged under the local coordinate transformation. In the second line, we made use of equation \eqref{eq:halooverdensity} and the long-wavelength part of $\delta_{\rm{mC}}$.
 To obtain the long-wavelength part of $\delta_{\rm{mC}}$ at second order, i.e $\delta\two_{{\rm mN} \l} $ and $\delta\two_{{\rm{mGR,C}} \l}$,  we first decompose equation \eqref{eq:GRdensity} into long and short-wavelength parts and transform to local coordinates. Long/short-wavelength coupling  terms in $\delta\two_{{\rm mN}} $ are absorbed into the local spatial curvature \cite{Dai:2015jaa}, {{the short/short-wavelength coupling terms in $\delta\two_{{\rm{mGR,C}}}$ are negligible as $\nabla^{-2} \delta\one_{\l} \gg \nabla^{-2} \delta\one_{s}$}}, while the long/short-wavelength coupling  terms in $\delta\two_{{\rm{mGR,C}}}$  are removed by local  coordinate transformation~(see equation \eqref{eq:densitycoord}),  leaving  
\begin{eqnarray}    \label{eq:longlongGRdensity}
\delta_{{\rm{mGR,C}} \l} \two(\tilde{\q})=6{ \Omega_{\rm m}\HH^2} \left[ -\frac{1}{4}\partial_i \nabla^{-2} \delta\one_{\l}(\tilde{\q}) \partial^i \nabla^{-2} \delta\one_{\l}(\tilde{\q})
    +\delta\one_{\l} (\tilde{\q})\nabla^{-2} \delta\one_{\l}(\tilde{\q}) \right] \Big(1+\frac{2}{3}{f\over\Omega_{\rm{m}}}\Big)\,.
\end{eqnarray}
This is how equation \eqref{eq:galaxibias} receives a general relativistic correction. Note that this contribution is not coming from the initial galaxy density, rather it is coming from the effect of volume distortions as the tracer evolves from $\tau_{\rm{ini}} $ to $\tau$. 
We  may now simplify equation \eqref{eq:galaxibias} further such that the second-order matter density contrast appears in the same form as the linear order term, i.e $\left[1 + b_1^{\rm{L}}\right]\delta_{\l}\one(\tilde{\q})$. A few steps of algebraic simplification lead to 
\begin{eqnarray} \label{eq:galaxyconservation2}
1+ \delta_{\!{\rm{gC}}\l}(\tilde{\q})& =&1 + \left[1 + b_1^{\rm{L}}\right]\delta_{\l}\one (\tilde{\q})
\\ \nonumber &&
+ \frac{1}{2} \bigg\{\left(1 + b_1^{\rm{L}}\right) \delta\two_{{\rm{m N}} \l}  (\tilde{\q})
 + \left[ b_{2}^{\text{L}} + \frac{2}{3}b^{\text{L}} _1\left(1-\frac{{F}}{{D}^2}\right) \right]\left(\delta\one_{\l}(\tilde{\q})\right)^2  + \left[  b_{s}^{\text{L}} -b_{1}^\text{L} \left(1-\frac{{F}}{{D}^2}\right)\right] s^2_{\l}(\q)+\delta_{{\rm{mGR,C}} \l}\two(\tilde{\q})\bigg\}\,.
\end{eqnarray}
%
We can now change $\tilde{q}$ to $q$ via coordinate transformation;
for $\delta_{{\rm{gC} }\l }$ we have
\begin{eqnarray}
\delta_{{\rm{gC}} \l }( {\q})  &=& \delta_{{\rm{gC}} \l} (\tilde{\q}) - \partial_j \delta_{{\rm{gC}}\l}\left({\q}\right)  \left(  \tilde{q}^j-{q}^j\right)\,.
\end{eqnarray}
There is a similar expression for $\delta_{\l}( {\q}) $.  Applying these to equation \eqref{eq:galaxyconservation2} and requiring that equation \eqref{eq:galaxyconservation2} is satisfied order by order, i.e $ \partial_j \delta_{{\rm{gC}} \l }\one =\left[1 + b_{1}^{\rm{L}}\right](\partial_{i}{\delta_{\l}\one})$, leads to
\begin{eqnarray} \label{eq:localbias}
\delta_{\!\rm{gC}}({\q})&=&\left[1 + b_1^{\rm{L}}\right]\left[ \delta_{\rm{m N}}\one(\q) + \frac{1}{2}\delta_{\rm{m N}}\two(\q) \right]
\\ \nonumber &&
 +\frac{1}{2}\left[ \left[ b_{2}^{\text{L}} + \frac{2}{3}b^{\text{L}} _1\left(1-\frac{{F}}{{D}^2}\right) \right]\left(\delta_{\rm{m}}\one(\q)\right)^2  +  \left[  b_{s}^{\text{L}} -b_{1}^\text{L} \left(1-\frac{{F}}{{D}^2}\right)\right] s^2(\q) +\delta_{{\rm{mGR,C}}}\two(\q) \right]\,.
\end{eqnarray}
We  omit the subscript $\l$ from now on as all perturbations in the above equation are long-mode and there is no confusion.


  \section{Eulerian gauges in global coordinates}
We have specified the galaxy density contrast in terms of the long-wavelength mode of the dark matter density contrast in C-gauge. As mentioned in  the introduction, in GR, there is no unique Eulerian frame. We can go to any convenient gauge by a performing coordinate transformation. In GR, perturbations change under a general coordinate transformation
\begin{eqnarray}
 {x}^{\mu} \rightarrow x^{\mu} + Z^{\mu}\,, \qquad {\rm{with}} \qquad  Z^{\mu} = ( T,L^i).
 \label{eq:gaugetransform}
\end{eqnarray}
Here $T$ stands for a temporal gauge choice and $L^i$ corresponds to spatial gauge choice and we decompose $L^i =\partial^i L+ L_{\bot}^i $, where $\partial _i L_{\bot}^i =0$.  
Consider a perturbed line element on an FLRW background  
\begin{eqnarray}
ds^2 = a^2 \Big[   - (1 + 2 \psi) d \eta^2 + 2 B_i dx^i d \eta + (1 - 2 \phi) \delta_{ij} dx^i dx^j 
+ 2 E_{ij} dx^i dx^j \Big]\,, 
\end{eqnarray}
 {where $B_i=\partial_iB$ and $E_{ij} =\big(\partial_i \partial_j - \frac{1}{3} \delta_{ij} \nabla^2 \big) E.$}  
 We consider two different  gauge choices, $X$-gauge and C-gauge. The gauge transformation  of the  density field from $X$ to C-gauge up to second order is given by  \cite{Bruni:1996im,Villa:2015ppa}
 \begin{eqnarray}\label{eq:Xgauge}
{\delta}_{ \rm{IX}}({\x})&=&\delta_{\rm{ IC}}({\x})- 3 {\cal H}T({\x})+\frac{1}{2}T({\x})\bigg[ 3\left( 2 {\cal H}^2 + \frac{3}{ 2} {\mathcal H^{2}\Omega_{{\rm m}}} \right)\,T({\x}) - 3 {\cal H}{T'({\x})}
\\ \nonumber &&
+2\delta_{\rm{IC}}'({\x}) - 6 {\cal H} \delta_{\rm{IC}}({\x}) \bigg]
+\left(\partial_i\delta_{ \rm{IC}}({\x}) - \frac{3}{2}{\cal H}\partial_i T({\x})\right)\partial^iL({\x})\,,
 \end{eqnarray}
where ${\rm{I}} = {\rm{g, m}}$ and $' = \partial_{\eta}$ and we omit the  $\eta$ dependence. Note that both ${\delta}_{ \rm{IC}}$ and $\delta_{ \rm{IX}}$ are calculated at the same coordinate position.
We consider the total matter gauge (T-gauge) which corresponds to an Eulerian gauge in the  Newtonian approximation and then give a general expression for any other Eulerian gauge choice.

\begin{itemize}
\item {\tt{Total matter gauge:}} In T-gauge, the components of the gauge vector are \cite{Villa:2015ppa}
\begin{eqnarray}\label{eq:gaugegenCT1}
L\one({\x})&=& \ \nabla^{-2}\delta_{\rm{m}}\one({\x}) \,, \qquad \qquad 
T = 0\,.\label{eq:gaugegenCT1B}
\end{eqnarray}
At  linear order the density does not change and at the second order, 
\begin{eqnarray}
\delta\two_{\rm{gT}}({\x}) &= &  \delta\two_{\rm{gC}}({\q})+2{\partial^j\nabla^{-2} \delta\one_{\rm{m}}}({\x}) \partial_j \delta\one_{\rm{gC}}({\x})\,,
\end{eqnarray}
 Putting everything together, the galaxy bias model in T-gauge becomes
\begin{eqnarray}\label{eq:Eulerianbias}
\delta_{\rm{gT}}({\x}) &=&
b_{1}\left[\delta\one_{\rm{m}}({\x}) + \frac{1}{2}\delta\two_{\rm{m N,T}}({\x})\right]
+\frac{1}{2}\bigg[b_{2}[\delta_{m}\one({\x})]^2
 +b_{s} s^2({\x}) +{\delta_{\rm{mGR,C}}\two({\x})} \bigg]\,,
\end{eqnarray}
 where we introduced the `Eulerian'  bias parameters
\begin{eqnarray}
b_{1}&=&1+b_{1}^\text{L}\,,\\
b_{2}&=&b_{2}^{\text{L}} + \frac{2}{3}\left(b_{1} -1\right)\left[1-\frac{{F}}{{D}^2}\right]=\frac{8}{21}\left(b_{1}-1\right)+b_{2}^\text{L}\,,\\
b_{s} &=& b_{s}^{\text{L}} -\left(b_{1} -1\right)\left[1-\frac{{F}}{{D}^2}\right]=b_{s}^\text{L}-\frac{4}{7}\left(b_{1} -1\right)\,.
\end{eqnarray}
The second equality holds in the Einstein de Sitter limit. 
$\delta\two_{\rm{ mN,T}}$ is the Newtonian limit of the second order matter density perturbation in T-gauge,
\begin{eqnarray}
\delta\two_{\rm{ mN,T}}({\x})&= &
2\partial_j \nabla^{-2}\delta_{\rm{m}}\one({\x}) \partial_j \delta_{\rm{m}}\one({\x})+ \frac{2}{3} \left(2+ \frac{F}{D^2}\right)  (\delta_{\rm{m}} \one({\x}))^2+  \left(1-\frac{F}{D^2}\right) s^2({\x})  \,.
\label{eq:TgaugeNewtonian}
\end{eqnarray}
This agrees with the conventional Newtonian result in the Eulerian frame. 

\item {\tt{X-Eulerian density:}}


Using  equation \eqref{eq:localbias} in equation \eqref{eq:Xgauge} gives
the general Eulerian galaxy density in $X-$gauge

\begin{eqnarray}\label{eq:GaldensityinX}
{\delta}_{\rm{gX}}({\x})&=&b_1\left[\delta\one_{\rm{mX}}({\x})+\frac{1}{2}\delta\two_{\rm{mX}}({\x})\right]+ \left(b_1-1\right) \left[{3}\HH T\one({\x})-\frac{1}{2}\left(\delta_{\rm{mGR,C}}\two({\x})- {3}\HH T\two({\x})\right)\right]
\\ \nonumber &&
+\frac{1}{2}\left[ b_2\left(\delta\one_{\rm{mX}}({\x})\right)^2  + b_{s}s^2({\x})\right] + \left( b_1' + 3 \HH b_2\right) \delta\one_{\rm{mX}}({\x})T\one({\x})
\\ \nonumber &&
+\frac{3}{2}\left(b_1 -1\right) \bigg[ \HH \partial_iT\one({\x}) \partial^i L\one ({\x})
- {T\one}'({\x})T\one({\x}) \bigg]
+\frac{3}{2}\bigg[\left(b_1-1\right) \left( 2 {\cal H}^2 
+ \frac{3}{ 2} {\mathcal H^{2}\Omega_{{\rm m}}} \right)\,
\\ \nonumber &&
 + b_1' \HH + 3b_2 \HH^2\bigg](T\one({\x}))^2 
\,.
\end{eqnarray}
Note that when $\delta\two_{\rm{mX}}({\x})$ is decomposed into Newtonian and GR corrections parts, $b_1\delta_{\rm{mGR,C}}\two({\x})$ drops out. 
The possible gauge choices include N-body-gauge for interpreting the Newtonian N-body simulation in the general relativistic context \cite{Fidler:2017pnb}  and  N-Boisson-gauge, which allows to  include the effects of radiation in the Newtonian N-body simulation \cite{Fidler:2018geb}.
\end{itemize}

{{Equation \eqref{eq:localbias}, with equation \eqref{eq:longlongGRdensity}, is our key result.}}


\section{Discussion and Conclusion}

It is well-known in cosmological perturbation theory that the comoving-synchronous  gauge completely fixes the gauge in a matter  {{plus cosmological constant}} dominated universe, {{once the initial matter world-line coordinates are fixed}}~\cite{Hwang:1999yv} .  So where does the extra freedom to re-define the coordinates  come from? It is important to stress that there is no contradiction with standard cosmological perturbation theory, since it assumes that the initial curvature perturbation ($\zeta = - 5\Phi^{\rm{ini}}/3$) falls off appropriately as $r \rightarrow \infty$~\cite{Hinterbichler:2012nm}. In our case, we made a key assumption that the galaxy formation process happens within a local patch of the  full spacetime. Based on this assumption, we split the initial curvature perturbation into long and short-wavelength modes, $\zeta = \zeta_s + \zeta_{\l}$, where short-wavelength refers to the modes that are equal or shorter than the size of the local patch. This allows us to express the initial metric  as 
\begin{eqnarray}
\gamma_{ij}^{\rm{ini}} = \delta_{ij} e^{2\zeta }= \delta_{ij} e^{2\zeta_{s}}e^{2\zeta_{\l}}.
\end{eqnarray}
The exponential conformal factor for the short-wavelength mode falls off in the limit $r\rightarrow R$.  
The exponential conformal factor for the long-wavelength mode does not die off in this limit, hence it is unobservable within the local patch.

 As a result it generates conformal transformations of the local patch which results in a residual diffeomorphism~\cite{Hinterbichler:2012nm,Weinberg:2003sw,Creminelli:2013mca}. We have  used this residual diffeomorphism symmetry to absorb $\zeta_{\l}$ into the coordinates of the FLRW background spacetime. If both $ \zeta_s$  and $\zeta_{\l}$ fall off as $r\rightarrow R$, there will not be any residual symmetry. Only in this limit can the comoving-synchronous gauge completely fix the gauge in the matter plus cosmological constant dominated universe. 
We used these residual diffeomorphism symmetries of GR to show that there is no coupling between $ \zeta_s$  and $\zeta_{\l}$ in the energy constraint equation in GR for a $\Lambda$CDM universe. This implies that there is no modulation of the local physics of galaxy formation by the long-wavelength mode of the metric perturbation.
{That is, at second order, no new terms appear in the bias expansion of equation~(\ref{eq:generalbiaspres}) when working in full GR rather than the Newtonian limit. The GR corrections only appear in the expressions for the operators in terms of, e.g. $\zeta$. This agrees with the conclusions reached by \cite{ip/schmidt}, who argued that new relativistic bias terms should only appear starting at third order.} 
We showed that the general relativistic effects do affect local clustering through the distortion of the volume element. 

{{In summary, our key result is \eqref{eq:localbias}, with equation \eqref{eq:longlongGRdensity}. We described in detail for the first time how to obtain a consistent expression for the local galaxy bias model at second order  in GR. We showed how the galaxy density is related to the underlying matter density field in both the Lagrangian frame (equation \eqref{eq:localbias}) and in the Eulerian frame (equation \eqref{eq:GaldensityinX}).  Our results show in a more transparent manner that the long wavelength mode associated with the GR corrections to the Poisson equation does not modulate galaxy clustering on small scales in a similar way that the local form of the primordial non-Gaussian does~\cite{Dalal:2007cu,Verde:2009hy}, rather; GR effects deform the volume element of the local patch as galaxies evolve.  }}


\acknowledgments

We thank Marco Bruni, Robert Crittenden and  David Wands for useful discussions.
{ Some of the tensor  algebraic computations here were done with the tensor algebra software xPand \cite{Pitrou:2013hga}}.
 OU, KK  and RM are supported by the UK STFC grant ST/N000668/1.  KK is also supported by the European Research Council under the European Union's Horizon 2020 programme (grant agreement 646702 ``CosTesGrav"). RM is also supported by the South African SKA Project and the National Research Foundation (Grant No. 75415). FS acknowledges support from the Starting Grant (ERC-2015- STG 678652) ``GrInflaGal'' from the European Research Council. CC   was   supported   by   STFC   Consolidated   Grant ST/P000592/1.

\appendix

\section{Perturbed metric in C-guage}\label{sec:CosmoPert}

The line element in C-gauge is given by
\begin{equation}
ds^2 = - d\tau^2 + a^2(\tau) \gamma_{i j}(\tau, \q) dq^i dq^j\,,
\end{equation}
where $\tau$ is the proper time and $a(\tau) $ is the scale factor.  The metric perturbation is  obtained by solving the Einstein field equation for dust plus cosmological constant domianted universe \cite{Villa:2015ppa} 
\begin{eqnarray}  \label{eq:evolvedgamma}
	 \gamma_{ij}(\tau,{\q}) &=& \gamma_{ ij}^{\rm{ini}}({\q})
    - \frac{4}{ 3} \frac{1}{\mathcal{H}^2\Omega_{\rm{ m}}} \partial_i \partial_j\Phi(\tau{,\q})- \frac{20}{9{\cal H}^2\Omega_{\rm{ m}}}  \partial_{i}  \Phi^{\rm{ini}}( {\q}) \partial_j 
    \Phi (\tau{,\q})
    \\ \nonumber &&
    +\frac{1}{2}\left[ 
   \frac{10}{9{\cal H}^2\Omega_{\rm m}} \partial_k\Phi^{\rm{ini}}( {\q}) \partial^k \Phi(\tau{,\q})\, - \frac{8}{9({\cal H}^2\Omega_{\rm m})^2}  \frac{ F}{D^2} \left([\nabla^2\Phi (\tau{,\q})]^2- \partial_l \partial_k\Phi(\tau{,\q}) \partial^l\partial^k \Phi (\tau{,\q})\right)\right] \delta_{ij} 
  \Bigg. \nonumber \\
 &&   - \frac{4}{9({\cal H}^2\Omega_{\rm m})^2} \left[ \partial_{i}\partial_k\Phi(\tau{,\q}) \partial^{k}\partial_j\Phi (\tau{,\q})-2 \frac{ F}{D^2} \left(
    2\partial_{i} \partial_{j}\Phi(\tau{,\q})\nabla^2\Phi(\tau{,\q}) -\partial_{i}\partial_{l}\Phi(\tau{,\q}) \partial_{j} \partial^l\Phi(\tau{,\q})\right)
  \right] \Bigg. \Bigg.\,.\nonumber 
\end{eqnarray}
$ \gamma_{ij}$ may also be decomposed in terms of the displacement field as  
\begin{eqnarray}
 \gamma_{ij}(\tau,{\q})  = \gamma_{  kl}^{\rm{ini}} ( {\q})\big[1 - 2 \mathcal{B}(\tau, {\q})  \big]  \big[ \delta^k_i +\partial_i\beta^k(\tau, {\q})  \big] \big[ \delta^l_j+\partial_j\beta^l(\tau, {\q})  \big] \,,
\end{eqnarray}
where $ \mathcal{B}$ is obtained from $ \gamma_{ij}$ \cite{Rampf:2014mga}.

\section{Comparison with conformal Fermi coordinates in C-gauge}\label{sec:CFC}


 {{
In conformal Fermi coordinates, the metric in the neighbourhood of the central geodesic is given by~\cite{Dai:2015rda}
\begin{eqnarray}\label{eq:CFCmetric}
g^F_{\mu\nu}(x^\mu_F) &=& a^2_F(\tau_F) \left[  \eta_{\mu\nu} + h^F_{\mu\nu} (\tau_F, x^i_F) \right],
\end{eqnarray}
where $h^F_{\mu\nu} $ is a small metric perturbation evaluated on the central geodesic, which may be  decomposed in terms of the  components of the Riemann tensor,
\begin{eqnarray}
h^F_{00} &=& -  R^F_{0l0m} x_F^l x_F^m\,, \\
h^F_{0i} &=& -\frac{2}{3} R^F_{0lim} x_F^l x_F^m \,,\\
h^F_{ij} &=& -\frac{1}{3}  R^F_{iljm} x_F^l x_F^m\,.
\end{eqnarray}
The local scale factor $a_F$  is constructed from the spacetime divergence of the 4-velocity of the central geodesic $H_F=\nabla_{\mu}u^{\mu}_F/3$,
\begin{eqnarray}
a_F(\tau_F)\propto \exp\left[\int \mathrm{d} \tau H_F\right]\,.
\end{eqnarray}
The relationship between the CFC coordinates and the global coordinates may be written as \cite{Dai:2015rda,Cabass:2016cgp}
\begin{eqnarray}\label{eq:CFCcoordinatemap}
x^\mu_F(\tau,\bm{x}) = x^\mu + \xi^\mu(\tau) + A^\mu_i(\tau)\, x^i
+ B^\mu_{ij}(\tau)\, x^i x^j + C^\mu_{kij}(\tau)\, x^i x^j x^k\,\,,
\end{eqnarray}
where $\xi^\mu,  A^\mu_i,  B^\mu_{ij}$ and $ C^\mu_{kij}$ are infinitesimally  small and  spatially constant. They are evaluated on the central geodesic. Note that equation \eqref{eq:CFCcoordinatemap} is expressed in terms of the CFC coordinates in \cite{Dai:2015rda,Cabass:2016cgp}, while we have expressed it in global coordinates since the coordinate transformation may be inverted order by order.  In C-gauge, equation \eqref{eq:CFCcoordinatemap} reduces to 
\begin{eqnarray}\label{eq:CFCcoordinatemapC-gauge}
x^j_F(\tau,{x}^i) = x^j + \xi^j(\tau) + A^j_i(\tau)\, x^i
+ B^j_{ij}(\tau)\, x^i x^j + C^j_{kij}(\tau)\, x^i x^j x^k\,\,,
\end{eqnarray}
where
\begin{eqnarray}
 \xi^j(\tau) &=& 0\,,\\ 
A^j_i(\tau) &=& -\left[(a_F/a)({\tau},\bm{0})- \zeta_0\right]\delta^j_i\,,\\ 
B^k_{ij}(\tau) &=& \frac{1}{2}\tilde{\Gamma}^k_{ij}(\tau,\bm{0})=\frac{1}{2}\left[ -\partial^k\zeta_0\delta_{ij} + \partial_i\zeta_0\delta^k_j + \partial_j\zeta_0\delta^k_i\right]\,\,,  \\
C^l_{kij}(\tau) &=& \frac{1}{6}\big[\partial_k\tilde{\Gamma}_{ij}^l({\tau},\bm{0}) - \mathcal{K}^l_{kij}({\tau},\bm{0})\big] =\frac{1}{6}\big[  -\partial_k\partial^l\zeta_0\delta_{ij} + \partial_k\partial_i\zeta_0\delta^l_{j} + \partial_k\partial_j\zeta_0\delta^l_{i}- \mathcal{K}^l_{kij}({\tau},\bm{0})\big]\,\,.
\label{eq:Cterm}
\end{eqnarray}
 $(a_F/a)({\tau},\bm{0})$ is the first order perturbation in the local scale factor evaluated on the central geodesic $[{\tau},\bm{0}]$ and $\zeta_0 \equiv \zeta(\tau,\bm{0})$ . The Christoffel connections are evaluated on the central geodesic. 
Note that some residual gauge freedom discussed in \cite{Cabass:2016cgp} has been used to introduce $\mathcal{K}^l_{kij}$ in  equation \eqref{eq:Cterm}. $\mathcal{K}^l_{kij}$ is related to the local curvature of the spatial section 
\begin{eqnarray}
\mathcal{K}^l_{kij}({\tau},\bm{0}) = -\frac{1}{6}K_F
(\delta^l_k\delta_{ij} + \delta^l_i\delta_{jk}+\delta^l_j\delta_{ki})\,, \qquad{\rm{with}} \qquad	K_F 
= -\frac{2}{3}\partial^2\zeta_0\,\,.
\end{eqnarray}
In the limit of vanishing local spatial curvature, $K_F=0$,  i.e vanishing double spatial derivatives of $\zeta$ on the central geodesic,  equation \eqref{eq:CFCcoordinatemapC-gauge} reduces to 
\begin{eqnarray}
x^i_F(\tau,\bm{x}) &=&x^i (1+ \hat{\zeta}_0)  + x^i {x}^j \partial_j \zeta_0  - \frac{1}{2}x_j x^j \partial^i \zeta_0\,\,,
\end{eqnarray}
where  we have set $ {\zeta}_{\bm{0}}-(a_F/a)({\tau},\bm{0}) \equiv \hat{\zeta}_{0}$. This is exactly the result we obtained from solving the conformal Killing equation in equation \eqref{eq:xcoordbody}.
}}

\section{Long mode and  density perturbations}\label{sec:generallongmode}
Consider  the short mode  component of the matter density perturbation in local coordinates
\begin{eqnarray}
\tilde{\rho}_s( \tilde{\q}) = \bar{\rho} +  \delta \rho_s ( \tilde{\q})\,,
\end{eqnarray}
where we omit the time dependence for brevity.
We can generate the long-wavelength mode by changing the coordinates  of $ \delta \rho_s ( \tilde{\q}) $, 
\begin{eqnarray}
\delta \rho_{{s}} ( \tilde{\q}) &=& \delta \rho_{{s}}({\q})  + \frac{\partial \delta \rho_{{s}}}{\partial \tilde{q}^j}\left({\q}\right)  \left(\tilde{q}^j - {q}^j\right) \,,
\\
&=&\delta \rho_{{s}}({\q})  +  \left[-  \frac{5}{3}q^i\Phi^{\rm{ini}}_{0}  -\frac{5}{3} q^i {q}^j (\partial_j \Phi^{\rm{ini}}_0) + \frac{5}{6} q_i q^i(\partial^i \Phi^{\rm{ini}}_0)\right]\frac{\partial \delta \rho_{{s}}\left({\q}\right)}{\partial \tilde{q}^j}  \,,
\end{eqnarray}
where we made use of equation \eqref{eq:xcoord2} in the second equality.
We simplify each term as follows:
\begin{itemize}
\item Dilatation: $-  \frac{5}{3}q^i\Phi^{\rm{ini}}_{0}  $
\begin{eqnarray}\label{eq:transformation}
{\delta}_{{s}} (\tilde{\q}) &=& \delta_{{s}}({\q}) -    \frac{5}{3}\Phi^{\rm{ini}}_{0}  \left( {q}^i \frac{\partial}{\partial {q}^i} \delta_{\rm{s}}({\q}) \right) \,,
\end{eqnarray}
where ${\delta}_{{s}} = \delta \rho_{{s}} /\bar{\rho}$. 
Expanding in Fourier space gives 
\begin{eqnarray}
 {\delta}_{{s}} (\tilde{\k}) = \delta_{{s}}({\k}) +\frac{5}{3}\Phi^{\rm{ini}}_{0} \left(3 +  \frac{d \log  {\delta}_{\rm{s}} ({\k}) }{d \log k}\right){\delta}_{{s}}  ({\k})\,.
\end{eqnarray} 
To simplify the term within the brackets, we use the linear order Poisson equation in C-gauge to relate to $\Phi$ in the Fourier space,
\begin{eqnarray}
\delta_{s}({\bm{k}}) = - \frac{2}{3\Omega_{\rm{m}}  \HH^2}    { k^2}\Phi({\bm{k}})\,.
\end{eqnarray}
We then relate $\Phi$ to the primordial potential $\varphi_{\bm{k}}$  through the transfer function  $T(k)$,  $\Phi(\tau,{\bm{k}})  =g(\tau)T(k)\varphi_{{\bm{k}}}$:  
\begin{equation}
3 + \frac{d {\rm{log}} {\delta_{\rm{s}}}({\bm{k}}) }{\d {\rm{log}} k}= 2 + 3 + \frac{d {\rm{log}}{\varphi}_{{\bm{k}}} }{d {\rm{log}} k}= 2+ \frac{d {\rm{log}} {T}(k) }{d {\rm{log}} k}
+ \frac{d {\rm{log}} (k^3\varphi_{\bm{k}})}{d {\rm{log}} k}\,.
\label{eq:dlogpar}
\end{equation}
The transfer function is constant on ultra-large scales $  T(k) \approx1$, and for a scale invariant initial power spectrum, the last term vanishes.  Putting everything  together leads to 
\begin{eqnarray}
{\delta}_{{s}} (\tilde{\q}) &=& \delta_{{s}}({\q})  +\frac{10}{3}\Phi^{\rm{ini}}_{0} \delta_{{s}}({\q}). 
\label{eq:transformdensity}
\end{eqnarray}
Using equation \eqref{eq:zetatodelta_L} leads to 
\begin{eqnarray}\label{eq:gaugeterm}
{\delta}_{{s}}(\tilde{\q}^i) & =&  \delta_{{s}}({\q})  
 + {3}{\Omega_{\rm{m}} \HH^2}\left( 1 + \frac{2 f}{3 \Omega_{\rm{m}}}   \right)\delta_{{s}}({\q}) \nabla^{-2} \delta_0\,.
\end{eqnarray}

\item The first term of the special conformal transformation: $ -\frac{5}{3} q^i {q}^j \partial_j \Phi^{\rm{ini}}_0$;

The response of the short mode to this part of the long-wavelength fluctuation is given by
\begin{eqnarray}
{\delta}_{{s}} (\tilde{\q}) &=& \delta_{{s}}( {\q})  -\frac{5}{3} \partial_j \Phi^{\rm{ini}}_0 \left(  q^i {q}^j  \frac{\partial}{\partial {q}^i} \delta_{{s}} ({\q})\right) \,.
\label{eq:specialconf1}
\end{eqnarray}
Expanding in Fourier space gives
\begin{eqnarray}
 {\delta}_{{s}} (\tilde{\k}) = \delta_{{s}}({\k}) + i \frac{5}{3} \partial_j \Phi^{\rm{ini}}_0  \left[ \frac{4}{\delta_{{s}}({\k})} \frac{\partial \delta_{{s}}({\k})}{\partial k_j}+ \frac{k_i}{\delta_{{s}}({\k})}\frac{\partial^2\delta_{{s}}({\k}) }{\partial k_i \partial k_j}\right]\delta_{{s}}({\k})\,.
\end{eqnarray}
We use the Poisson equation again for the term within the square bracket,
\begin{eqnarray}
\frac{1}{\delta_{{s}}({\bm{k}})} \frac{\partial \delta_{{s}}({\k})}{\partial k_j}& =& \frac{1}{\Phi}\frac{\partial \Phi}{\partial k_j} + 2 \frac{k^j }{k^2}\,,
\\
\frac{1}{\delta_{{s}}{(\bm{k}})}  \frac{\partial^2\delta_{\rm{s}}({\k}) }{\partial k_i \partial k_j} &=& \frac{1}{\Phi}\frac{\partial^2 \Phi}{\partial k_i \partial k_j} + 2\left[\frac{k^i }{k^2}\frac{1}{\Phi}\frac{\partial \Phi}{\partial k_j} + \frac{k^j }{k^2}\frac{1}{\Phi}\frac{\partial \Phi}{\partial k_i}\right] + \frac{2}{k^2} \delta^{ij} \,.
\end{eqnarray}
On large scales, for the   initial scale invariant  potential,
\begin{eqnarray}
\frac{1}{\varphi_{{\bm{k}}}} \frac{\partial \varphi_{{\bm{k}}}}{\partial k_i} &=& \frac{k^i}{k^2} \bigg[  \frac{d\log (k^3\varphi_{{\bm{k}}})}{d\log k} -3\bigg] \approx-3\frac{k^i}{k^2} \,,
\\
\frac{k^i}{\varphi_{{\bm{k}}}}\frac{\partial^2\varphi_{{\bm{k}}}}{\partial k_i \partial k_j} &=&  \frac{k^j}{k^2}\bigg[    \frac{d^2\log (k^3\varphi_{{\bm{k}}})}{d(\log k)^2}   + \left[\frac{d\log (k^3\varphi_{{\bm{k}}})}{d\log k} \right]^2 - 7 \frac{d\log (k^3\varphi_{{\bf{k}}})}{d\log k}  + 12\bigg]\approx 12 \frac{k^j}{k^2}\,.
\end{eqnarray}
Upon simplification this  leads to 
\begin{eqnarray}
 {\delta}_{{s}} ( \tilde{\k})  &=&\delta_{{s}}( {\k}) -\frac{10}{3} \partial_j \Phi^{\rm{ini}}_0\left[ \frac{k^j_s}{k^2_s}\delta_{{s}}({\k})\right]\,.
\end{eqnarray}
Remapping back to real space, using  $-k^2_s = \nabla^2$, $ i k^j_s= \partial^j$,  gives
\begin{eqnarray}
 {\delta}_{{s}} ( \tilde{\q}) &=&
 \delta_{{s}}({\q}) + {3} \Omega_{\rm{m}} \HH^2\left(1+ \frac{2f}{3 \Omega_{\rm{m}}} \right) \partial^j\nabla^{-2} \delta_{{s}}({\q})\partial_j  \nabla^{-2}\delta_{0}\,.
\end{eqnarray}

\item The second part of the special conformal transformation $\frac{5}{6} q_i q^i\partial^i \Phi^{\rm{ini}}_0 $:

The coordinate transformation involving this term is given by
\begin{eqnarray}
{\delta}_{{s}} ( \tilde{\q}) &=& \delta_{{s}}( {\q}) +  \frac{5}{6} \partial^i \Phi^{\rm{ini}} _0\left(  q_j {q}^j \partial_i \delta_{{s}} ({\q})\right) \,.
\end{eqnarray}
Evaluating the derivative in Fourier space leads to 
\begin{eqnarray}
{\delta}_{{s}} (\tilde{\k}) &=& \delta_{{s}}({\k}) -  \frac{5}{6} \partial^i \Phi^{\rm{ini}}_0 \left[ \frac{2}{\delta_{\rm{s}}{(\bm{k}})} \frac{\partial \delta_{{s}}({\k})}{\partial k_i}+ \frac{k^i}{\delta_{{s}}{(\bm{k}})}\frac{\partial^2\delta_{{s}}({\k}) }{\partial k^2}\right]\delta_{{s}}({\k})
\,.
\label{eq:specialconf2} 
\end{eqnarray}
The simplification of equation \eqref{eq:specialconf2} is equivalent to the simplification of equation \eqref{eq:specialconf1}, so a similar procedure applies,
\begin{eqnarray}
{\delta}_{{s}} ( \tilde{\k})
&=& \delta_{{s}}({\k}) +  i \frac{5}{3} \partial_j \Phi^{\rm{ini}}_0\frac{k^j}{k^2} \delta_{{s}}({\k})\,.
\end{eqnarray}
Again using $-k^2_s = \nabla^2$, $ i k^j_s= \partial^j$  and equation \eqref{eq:zetatodelta_L} leads to 
\begin{eqnarray}
{\delta}_{{s}} ( \tilde{\q})&=& \delta_{{s}}({\q})  - \frac{3}{2} \Omega_{\rm{m}} \HH^2\left(1+ \frac{2f}{3 \Omega_{\rm{m}}} \right)\partial^j\nabla^{-2} \delta_{{s}}({\q})\partial_j  \nabla^{-2}\delta_{0}\,.
\end{eqnarray}

\end{itemize}

Bringing all the terms together leads to
\begin{eqnarray}
{\delta}_{{s}}(\tilde{\q}) & =&  \delta_{{s}}({\q})  
+{3}{\Omega_{\rm{m}} \HH^2}\left( 1 + \frac{2 f}{3 \Omega_{\rm{m}}}   \right)\delta_{{s}}({\q}) \nabla^{-2} \delta_{0}
+ \frac{3}{2} \Omega_{\rm{m}} \HH^2\left(1+ \frac{2f}{3 \Omega_{\rm{m}}} \right) \partial^j\nabla^{-2} \delta_{{s}}({\q})\partial_j  \nabla^{-2}\delta_{0}\,.
 \label{eq:gaugetermgaussian}
\end{eqnarray}


\providecommand{\href}[2]{#2}\begingroup\raggedright\endgroup

\end{document}